\documentclass[twocolumn,english,pra,preprintnumbers,showpacs,nofootinbib]{revtex4}
\usepackage{graphicx}
\usepackage{babel}
\usepackage{slashed}

\newcommand{\lp}{\left(}
\newcommand{\rp}{\right)}
\newcommand{\nn}{\nonumber}
\newcommand{\al}{\alpha}
\newcommand{\be}{\begin{equation}}
\newcommand{\ee}{\end{equation}}
\newcommand{\bea}{\begin{eqnarray}}
\newcommand{\eea}{\end{eqnarray}}

\begin{document}

\title{\boldmath Radiative-nonrecoil corrections of order $\al^{2}(Z\al)E_F$ to the hyperfine splitting of muonium }

\preprint{ALBERTA-THY-06-10}
\preprint{CERN-PH-2010-099}

\author{Jorge Mond{\'e}jar}
\author{Jan H. Piclum}
\affiliation{Department of Physics, University of Alberta, Edmonton, Alberta,
Canada T6G 2G7}
\author{Andrzej Czarnecki}
\affiliation{Department of Physics, University of Alberta, Edmonton, Alberta,
Canada T6G 2G7 and CERN Theory Division, CH-1211 Geneva 23, Switzerland}

\date{\today}
\begin{abstract}
We present results for the corrections of order $\al^{2}(Z\al)E_F$
to the hyperfine splitting of muonium. We compute all the contributing Feynman diagrams
in dimensional regularization and a general covariant gauge using
a mixture of analytical and numerical methods. We improve the precision of previous results.
\end{abstract}


\maketitle

\section{Introduction}

Muonium is the hydrogenlike bound state of a positive muon and an electron. Unlike hydrogen, or any 
other bound state involving hadrons, muonium is free from the complications introduced by the finite 
size or the internal structure of any of its constituents.
Therefore, it allows for a very precise test of bound-state QED, and can be used to restrict models 
of physics beyond the Standard Model. Measurements of the ground-state hyperfine splitting of muonium 
are used to extract the muon to electron mass ratio $m_{\mu}/m_e$ and the muon to proton magnetic moment
ratio
$\mu_{\mu}/\mu_{p}$ \cite{Mohr:2008fa}. The value of $\mu_{\mu}/\mu_{p}$ is required for obtaining the muon
anomalous magnetic moment from experiment \cite{Roberts:2010cj}. In addition, the hyperfine splitting can 
also be used to determine the fine structure constant $\alpha$.
For a review
of the present status and recent developments in the theory of light
hydrogenic atoms, see~\cite{Eides:2000xc, book}.

The leading-order hyperfine splitting is given by the Fermi energy $E_F$ (defined in Eq. (\ref{EF})). Its corrections
are organized as a perturbative expansion in powers of three parameters: $Z\alpha$,
describing
effects due to the binding of an electron to a nucleus of atomic number
$Z$; $\alpha$ (frequently accompanied by $1/\pi$) from electron and photon
self-interactions; and the ratio of electron to nucleus masses, $m/M$.
The main theoretical uncertainty 
comes from three types of yet unknown corrections: single-logarithmic and non-logarithmic corrections of order $\alpha(Z\alpha)^2(m/M)E_F$,  
and non-logarithmic corrections of order $\alpha^2(Z\alpha)(m/M)E_F$ and $(Z\alpha)^3(m/M)E_F$ (some 
terms are known for the first case \cite{Eides:2009dp}).

In this paper we focus on the second-order radiative-nonrecoil corrections to the hyperfine splitting (of 
order $\al^{2}(Z\al)E_F$). The total result for these corrections was found by Eides and Shelyuto \cite{Eides:1995ey} 
and Kinoshita and Nio \cite{Kinoshita:1995mt}. Our result improves their precision by over an order of magnitude.
Our central value is slightly lower than, but compatible with, that of \cite{Eides:1995ey}.

In Sec. \ref{secEval} we present the details of our approach, and in Sec. \ref{secRes} we present our results.
In Appendix \ref{appA} we show analytic results for two sets of diagrams.

\section{Evaluation}
\label{secEval}

We consider an electron of mass $m$ orbiting a nucleus
of mass $M$ and atomic number $Z$.
In this paper we consider the nucleus to be a muon, but we will keep $Z$
explicit in order to distinguish between the binding contributions ($Z\alpha$) and the radiative ones ($\alpha$).

\begin{figure}[t]
  \includegraphics[width=\columnwidth]{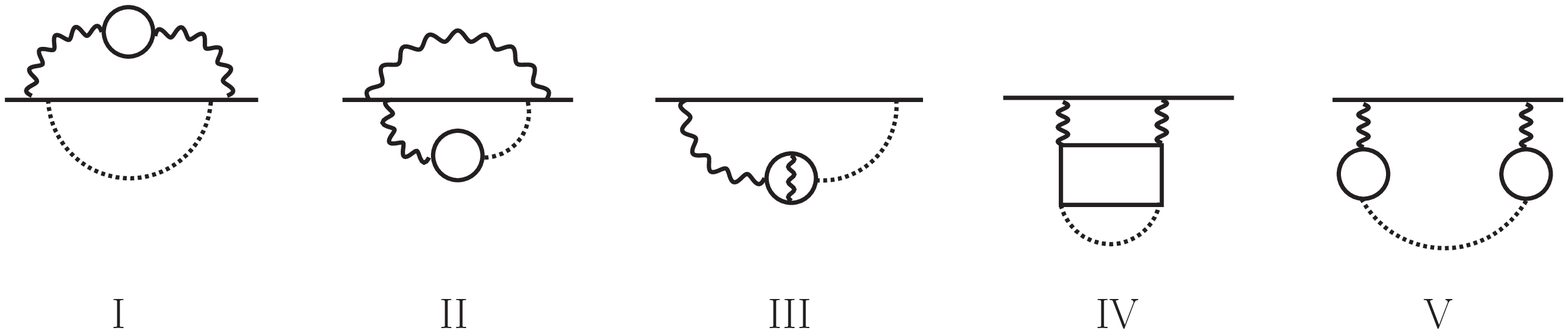}
  \caption{\label{vacuum}The different sets of vacuum polarization
      diagrams and light-by-light diagrams (set IV). Each set represents the drawn diagram plus all the
      possible permutations of its pieces.}
\end{figure}

\begin{figure}[t]
  \includegraphics[width=\columnwidth]{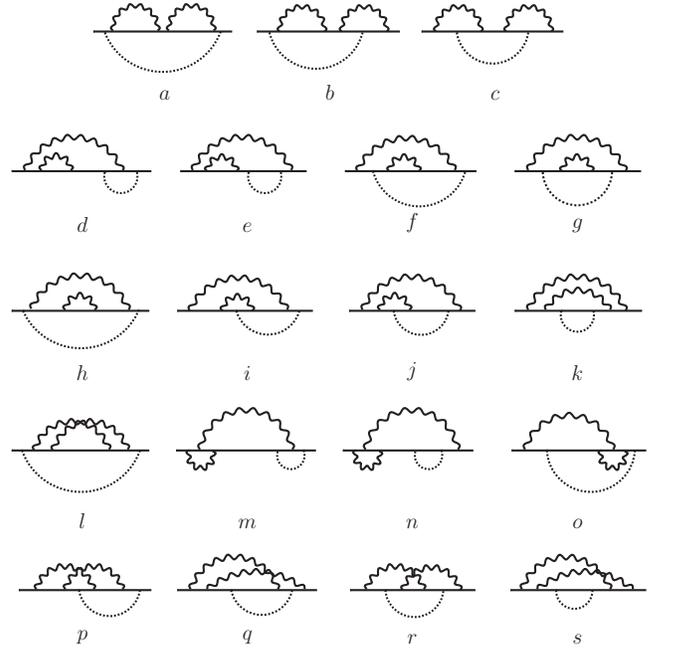}
  \caption{\label{other}The diagrams involving a two-loop electron
    self-interaction and vertex corrections.}
\end{figure}

\begin{figure}[t]
  \includegraphics[width=\columnwidth]{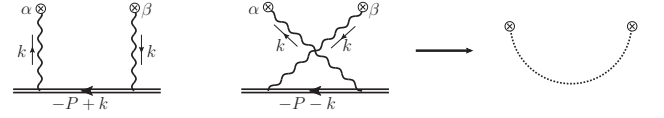}
  \caption{\label{graphdelta}The sum of the direct and crossed diagrams
    is represented by a dotted line (the double line
    represents the propagator of the muon).}
\end{figure}

We are
interested in corrections to the hyperfine splitting of the ground state of muonium of order $ \al^{2}(Z\al)E_F$ and
leading order in $m/M$, where
\be
\label{EF}
E_F=\frac{8}{3}\frac{\mu^3(Z\alpha)^4}{m M}\frac{g}{2}\,.
\ee
Here $g$ is the  gyromagnetic factor of the nucleus\footnote{It includes the corrections from the anomalous magnetic
moment, which factorize with respect to the corrections considered in this paper. This is no longer true when 
considering non-recoil corrections. See e.g. \cite{Eides:2000xc}.} (in our case, a muon, but our final result in Eq.~(\ref{res}) 
applies to any hydrogenlike atom).
In order to compute these corrections, we consider the scattering amplitude
\be
\label{M1}
i\mathcal{M}=[\bar{u}(p)\mathcal{Q}_1u(p)][\bar{v}(P)\mathcal{Q}_2v(P)] \,,
\ee
where $u(p)$ is the spinor for the electron, $v(P)$ is the spinor for the muon, $p=(m,\vec{0})$
and $P=(M,\vec{0})$. $\mathcal{Q}_1$ and $\mathcal{Q}_2$ are given by the Feynman rules describing the sum of the
diagrams shown in
Figs.~\ref{vacuum} and \ref{other}. In these figures, the sum of the direct and crossed interactions between the 
electron and the muon is
represented by a dotted line, as shown in Fig.~\ref{graphdelta}.
We define a bound-state wave function $\psi=u\bar{v}$, so that Eq.~(\ref{M1}) becomes
\be
i\mathcal{M}=-\text{Tr}\{\psi^{\dagger}\mathcal{Q}_1\psi\mathcal{Q}_2\} \,.
\ee

Depending on the relative alignment of the spins of the constituent particles, an $S$ state can either belong to 
the $J=1$ triplet or the $J=0$ singlet. The triplet and singlet states are often denoted by the prefixes ortho-
and para-, respectively, and their wave functions are given by \cite{Czarnecki:1998zv}
\bea
\psi_{para}&=&\frac{1+\gamma_0}{2\sqrt{2}}\gamma_5 \,,\\
\psi_{ortho}&=&\frac{1+\gamma_0}{2\sqrt{2}}\vec{\gamma}\cdot\vec{\xi}\,,
\eea
where $\vec{\xi}$ is the polarization vector. We average over the directions of $\vec{\xi}$ by considering 
the four-vector $\xi\equiv(0,\vec{\xi})$ and using the
identity
\be
\langle (\xi\cdot A)(\xi\cdot B)\rangle=\frac{1}{d-1}\lp A_0B_0-A\cdot B\rp \,.
\ee
We use dimensional regularization with $d=4-2\epsilon$ dimensions. Thus, an important issue is the definition
of $\gamma_5$, which is an intrinsically four-dimensional object. Since we do not have to evaluate traces
with an odd number of $\gamma_5$ matrices, we can treat them as anticommuting.

The energy shift created by the radiative corrections depicted in Figs.~\ref{vacuum} and \ref{other}, for either
the singlet or triplet configurations, is given by
\begin{equation}
  \delta E=-|\psi_{n}(0)|^{2}\mathcal{M}\,,
\end{equation}
where $|\psi_{n}(0)|^{2}=(Z\al\mu)^{3}/(\pi n^{3})$ is the squared
modulus of the wave function of a bound $S$ state with principal
quantum number $n$  and reduced mass $\mu$.
The hyperfine splitting is then simply
\be
\delta E_{hfs}=\delta E_{ortho} - \delta E_{para} \,.
\ee

In order to evaluate the loop integrals represented by the Feynman diagrams we use the method of regions~\cite{Smirnov:2002pj} to construct an
expansion in the small ratio $m/M$. There
are several possible contributing
regions, where one or more of the loop momenta scale like $m$ or $M$. However, we are only interested in the leading
order in $m/M$, which is given by the region where all loop momenta scale like $m$.
 If $k\sim m$, we can expand the contribution from the muon line in the sum of the direct and crossed diagrams of
 Fig.~\ref{graphdelta},
\bea
\label{exp}
&&\gamma_{\alpha} \frac{\slashed{k}-\slashed{P}+M}{(k-P)^2-M^2+i\epsilon}\gamma_{\beta} - \gamma_{\beta} \frac{\slashed{k}
+\slashed{P}-M}{(k+P)^2-M^2+i\epsilon}\gamma_{\alpha}\nn\\
&& \to T_1+T_2+T_3\,,
\eea
where
\bea
T_1=&
  2P_{\beta}\gamma_{\alpha}\left[\lp \frac{1}{2P\cdot k-i\epsilon} -  \frac{1}{2P\cdot k+i\epsilon}\rp+ \mathcal{O}\lp\frac{1}{P^2}\rp\right]\,, \\
T_2 =&-\gamma_{\alpha}\slashed{k}\gamma_{\beta}\left[\lp \frac{1}{2P\cdot k-i\epsilon} -  \frac{1}{2P\cdot k+i\epsilon}\rp
 + \mathcal{O}\lp\frac{1}{P^2}\rp\right]\,,\\
 T_3=&-\lp \gamma_{\alpha}\slashed{k}\gamma_{\beta} + \gamma_{\beta}\slashed{k}\gamma_{\alpha}\rp \left[\frac{1}{2P\cdot k +i\epsilon}
 + \mathcal{O}\lp\frac{1}{P^2}\rp\right] \,.
   \eea
We used the equation of motion to set some terms in the numerator to zero, and we arranged the terms in the expansion in such a way that the three different Dirac structures that are important for the calculation of the hyperfine splitting appear explicitly. We will now see that only $T_2$ can contribute to the splitting.
 
Consider the Dirac structure of $\psi T_1$ and anticommute the gamma matrices, for both para and ortho states:
 \bea
 \label{para1}
 \chi^{para}_{T_1}&\equiv&\frac{1+\gamma_0}{2\sqrt{2}}\gamma_5\,\gamma_{\alpha}\nn\\
 &=&-\gamma_{\alpha}\frac{1-\gamma_0}{2\sqrt{2}}\gamma_5
  - \frac{1}{\sqrt{2}}g_{\alpha 0}\gamma_5 \,,\\
  \label{ortho1}
\chi^{ortho}_{T_1}&\equiv& \frac{1+\gamma_0}{2\sqrt{2}}\gamma_i\,\gamma_{\alpha}\nn\\
&=&-\gamma_{\alpha}\frac{1-\gamma_0}{2\sqrt{2}}\gamma_i - \frac{1}{\sqrt{2}}
 g_{\alpha 0}\gamma_i+\frac{1}{\sqrt{2}}g_{\alpha i}(1+\gamma_0)\,.\nn\\
\eea
Now we can write
\bea
i\mathcal{M}_{T_1}&\equiv& -\text{Tr}\{\psi^{\dagger}\mathcal{Q}_1\psi T_1\}\nn\\
&\propto& \text{Tr}\{\psi^{\dagger}\mathcal{Q}_1\chi_{T_1}\}=
\text{Tr}\{\chi_{T_1}\psi^{\dagger}\mathcal{Q}_1\} \,.
\eea
Using the expressions in Eqs. (\ref{para1}) and (\ref{ortho1}) it is easy to see that $\chi^{para}_{T_1}\psi_{para}^{\dagger}=\chi^{ortho}_{T_1}\psi_{ortho}^{\dagger}$ (after averaging
over polarizations). This means that $T_1$ gives the same contribution for
para and ortho states.
Therefore, when we subtract these contributions in order to compute the hyperfine splitting, they cancel out.

If we consider $T_2$ instead, defining $\chi_{T_2}$ in analogy with Eqs.~(\ref{para1}) and (\ref{ortho1}), we can see that  $\chi^{para}_{T_2}\psi_{para}^{\dagger}
\neq\chi^{ortho}_{T_2}\psi_{ortho}^{\dagger}$, so this term will not cancel in the subtraction. The difference
between the para and ortho states comes solely from terms in $\chi^{ortho}_{T_2}$ that are totally antisymmetric in
$\alpha$ and $\beta$. Therefore, when we consider the Dirac structure of $T_3$, which is but a symmetrization
of that of $T_2$, these terms will vanish, and so $T_3$ will give
no contribution to the hyperfine splitting either.

Thus, we have seen that the only term that contributes to the hyperfine splitting is
\be
\label{split}
-\gamma_{\alpha}\slashed{k}\gamma_{\beta}\left[\lp \frac{1}{2P\cdot k-i\epsilon} -  \frac{1}{2P\cdot k+i\epsilon}\rp + \mathcal{O}\lp\frac{1}{P^2}\rp\right] \,.
\ee
This is valid at all orders of alpha, and all orders in $m/M$.
We can then substitute the scalar part of the nucleon propagator by a Dirac delta in all our calculations, since
we are only interested in the leading order in $m/M$ and
\be
 \frac{1}{2P\cdot k-i\epsilon} -  \frac{1}{2P\cdot k+i\epsilon} = i\pi\delta\lp P\cdot k\rp \,.
 \ee

We used dimensional regularization, and renormalized our results using
the on-shell renormalization scheme. For all the photon propagators
in Figs.~\ref{vacuum} and \ref{other} we used a general covariant
$R_{\xi}$ gauge. The overall cancellation of the dependence on the
gauge parameter in the final result provides us with a good check
for our calculations.

We used the program \texttt{qgraf}~\cite{Nogueira:1991ex} to generate
all of the diagrams, and the packages \texttt{q2e} and
\texttt{exp}~\cite{Harlander:1997zb,Seidensticker:1999bb}
to express them as a series of vertices and propagators that can be
read by the \texttt{FORM}~\cite{Vermaseren:2000nd} package \texttt{MATAD~3}~\cite{Steinhauser:2000ry}.
Finally, \texttt{MATAD~3} was used to represent the diagrams in terms
of a set of scalar integrals using custom-made routines. In this way, we
represented the amplitude $\mathcal{M}$ in terms of several thousand different scalar integrals. These integrals
can be expressed
in terms of a few master integrals by means of integration-by-parts
(IBP) identities~\cite{Chetyrkin:1981qh}. We used the so-called Laporta
algorithm \cite{Laporta:1996mq,Laporta:2001dd} as implemented in
the \texttt{Mathematica} package \texttt{FIRE}~\cite{Smirnov:2008iw},
to reduce the problem to 32 master integrals. The master integrals for this calculation are the same ones
we found in \cite{Dowling:2009md}. All definitions and results for the integrals can be found in 
this reference. However, 
one change was made for this calculation. In order to obtain better numerical precision,
we performed a change of basis, so that instead of working with $I_{14}=F(1,0,0,0,1,1,1,1)$ we worked with
\bea
\lefteqn{F(1,0,0,0,1,1,1,2)}&&\nn \\
&=&44.55822275(2) - 427.382296(2)\epsilon+ \mathcal{O}(\epsilon^2) \,,
\eea
which was obtained using the \texttt{Mathematica} package \texttt{FIESTA~1.2.1}~\cite{Smirnov:2008py}
with integrators from the \texttt{CUBA} library~\cite{Hahn:2004fe}.

\section{Results}
\label{secRes}

Our final result for the hyperfine splitting is
\be
\label{res}
\delta E_{hfs}=0.77099(2)\cdot \frac{\alpha^2(Z\alpha)}{\pi n^3}E_F  \,.
\ee
This correction was also found by Eides and Shelyuto~\cite{Eides:1995ey}, and by Kinoshita and 
Nio~\cite{Kinoshita:1995mt}. Their results are
\bea
\delta E_{hfs}&=&0.7716(4)\cdot\frac{\alpha^2(Z\alpha)}{\pi n^3}E_F\quad\,\,\,\text{\cite{Eides:1995ey}} \,,\\ 
\delta E_{hfs}&=&0.7679(79)\cdot\frac{\alpha^2(Z\alpha)}{\pi n^3}E_F\quad\text{\cite{Kinoshita:1995mt}}\,.
\eea
Our result is a little over one order of magnitude more precise than that of \cite{Eides:1995ey}, and almost three orders
of magnitude more precise than the one in \cite{Kinoshita:1995mt}. Our central value is slightly lower than in 
\cite{Eides:1995ey}, by about $1.5\sigma$ (taking as $\sigma$ the larger error). It agrees with the result of 
\cite{Kinoshita:1995mt} within its much larger error estimate.
For the ground state of muonium, our result reads
\be
\delta E_{hfs} = 0.42524(1)\, \text{kHz} \,.
\ee

We compared our results for the individual
diagrams and those found in the literature~\cite{EKS2,EKS2b,EKS3,Eides:1995ey,Kinoshita:1995mt}.
Our results for the gauge-invariant sets of diagrams of Fig.~\ref{vacuum} are presented in Table~\ref{tvac}.
For the diagrams of Fig.~\ref{other} we chose the Fried-Yennie gauge \cite{Fried:1958zz,Adkins:1993qm}, in which all diagrams
are infrared finite. Our results are presented in Table~\ref{tother}.

\begin{table}[tb]
      \caption{\label{tvac} Comparison between our results for sets of diagrams of Fig. \ref{vacuum} and those of
    \cite{EKS2,EKS2b,EKS3}. Numbers ending in an ellipsis
    indicate an analytic result, which we show in Appendix \ref{appA}. No error was given for the numerical result of set I in \cite{EKS2b}.}
    \begin{ruledtabular}
    \begin{tabular}{lll}
      Set  & This paper  & Refs. \cite{EKS2,EKS2b,EKS3}\\
      \hline
      I&$ -0.31074204276602(3) $&$ -0.310742$\\
       II&$ -0.668915\dots $&$ -0.668915\dots$\\
       III&$1.867852\dots  $&$ 1.867852\dots$\\
       IV&$ -0.4725146(2) $&$ -0.472514(1)$\\
       V&$ 36/35 $&$ 36/35$\\
    \end{tabular}
  \end{ruledtabular} 
\end{table}

\begin{table}[t]
  \caption{\label{tother} Comparison between our results for diagrams
    $a$--$s$ (in Fried-Yennie gauge) and those of
    \cite{Eides:1995ey}.}
  \begin{ruledtabular}
    \begin{tabular}{lll}
      Diagram  & This paper  & Ref. \cite{Eides:1995ey}\\
      \hline
     $a$ & $9/4$ & $9/4$\\
      $b$ & $-6.6602948853575169751(3)$ & $-6.65997(1)$\\
      $c$ & $3.9324055550472089860(4)$ &$ 3.93208(1)$\\
      $d$ & $-3.9032816968990(2)$ & $-3.903368(79)$\\
      $e$ &$4.5667195410288(2)$ & $4.566710(24)$\\
      $f$ & $-3\pi^2/8+19/64$ & $-3.404163(22)$\\
      $g$ & $\pi^2/2-9/4$ & $2.684706(26)$\\
      $h$ & $33/16$ & $33/16$\\
      $i$ & $0.05454(1)$ & $0.054645(46)$\\
      $j$ & $-7.14963(2)$ & $-7.14937(16)$\\
      $k$ &$1.4658690989997(5)$  & $1.465834(20)$\\
      $l$ &$-1.98334(3)$  & $-1.983298(95)$\\
      $m$ & $3.16949(2)$ & $3.16956(16)$\\
      $n$ & $-3.59661163(2)$ &$ -3.59566(14)$\\
      $o$ & $1.80476(5)$ & $1.804775(46)$\\
      $p$ & $3.507035(6)$ &$ 3.50608(16)$\\
      $q$ & $-0.80380(3)$ &$ -0.80380(15)$\\
      $r$ & $1.05247 (3)$& $1.05298(18)$\\
      $s$ & $0.277336777308(2)$ & $0.277203(27)$\\
          \end{tabular}
  \end{ruledtabular} 
\end{table}

The sum of all central values in the second column of Tables~\ref{tvac} and \ref{tother} gives the coefficient $0.77099$ in Eq.~(\ref{res}).
The error of that result is however not obtained from the sum of the
errors of the diagrams in the tables. Once
we decompose the problem into the calculation of master integrals,
the diagrams are no longer independent, as the same master integral
contributes to several different diagrams. Thus, to find the error
of our total result, we first sum all diagrams and then sum all the
errors of the integrals in quadrature.

We found new analytic results for diagrams $g$ and $f$, shown in Table~\ref{tother}. For completeness, the known analytic results
for sets II and III of the vacuum polarization diagrams are given
in Appendix~\ref{appA} as well.

We found no discrepancies between our results for the diagrams of Fig.~\ref{vacuum} and the ones of 
\cite{EKS2,EKS2b,EKS3}, but we found significant differences in the rest of the diagrams between the results
of \cite{Eides:1995ey} and ours. They affect all diagrams except
diagrams $a$, $e$, $h$, $l$, $o$, and $q$. The biggest 
discrepancies are in diagrams $b$ and $c$, and they are of the order of $30\sigma$.
However, most of the differences cancel when summing the diagrams. In particular, there are almost exact
cancellations between the differences in diagrams $b$ and $c$, $k$ and $l$, and $n$ and $p$.

The reason for the discrepancies (and their cancellations) is most likely the different treatment of infrared divergences in \cite{Eides:1995ey}
and this paper. In \cite{Eides:1995ey}, the Fried-Yennie gauge was set from the beginning, and all spurious infrared divergences were
canceled before the integration over the diagram's loop momenta, which was performed in four dimensions. In our calculation, we used a general gauge parameter, and dimensional
regularization to deal with infrared divergences, which would only vanish after setting the gauge in the final expression. 
As noted in \cite{Tomozawa:1979xn}, there is a difference between setting the Fried-Yennie gauge and sending the infrared regulator to zero
before or after integration. It is not surprising then that we obtained different results than \cite{Eides:1995ey} for gauge-dependent diagrams, but that 
most of the differences cancel in the final, gauge-invariant result, making it compatible with the previous calculation.

Using the setup of the calculation of the hyperfine splitting one can also find the Lamb shift, as it is given by
\be
\label{defLamb}
\delta E_{Lamb}=\frac{\delta E_{ortho}\,(d-1)+\delta E_{para}}{d} \,.
\ee
We obtained in this way the same results as in \cite{Dowling:2009md}\footnote{There is a mistake in the values in 
the last row of Table I in the published version of \cite{Dowling:2009md} (it was corrected in version 3 of the preprint on the arXiv). They 
read $-23/278$, when they should be $-23/378$. This does not affect any of the other results presented in that paper.}.

\begin{acknowledgments}
We thank M.~I.~Eides for
helpful comments. This work was supported by the Natural Sciences and
Engineering Research Council of Canada. The work of J.H.P. was supported
by the Alberta Ingenuity Foundation. The Feynman diagrams were drawn
using \texttt{Axodraw}~\cite{Vermaseren:1994je} and
\texttt{Jaxodraw 2}~\cite{Binosi:2008ig}.
\end{acknowledgments}

\appendix

\section{Analytic results}
\label{appA}

Here we show the analytic results for sets II and III of the vacuum polarization diagrams, found in \cite{EKS2},
\begin{eqnarray}
  \text{Set II} & = &-\frac{4}{3}\ln^2\lp \frac{1+\sqrt{5}}{2}\rp -\frac{20}{9}\sqrt{5}\,\text{ln}\lp\frac{1+\sqrt{5}}{2}\rp \nn\\
 && -\frac{64}{45}\ln2+\frac{\pi^2}{9}+\frac{10369}{5400}\,, \\
 \nn \\
  \text{Set III} & = &\frac{224}{15}\ln2 - \frac{38}{15}\pi-\frac{118}{225}\,.
\end{eqnarray}

\end{document}